\documentclass[twocolumn,showpacs,preprintnumbers,amsmath,amssymb]{revtex4-1}
\usepackage{bbm}
\usepackage{amsfonts}

\usepackage{epsfig}
\usepackage{graphicx}% Include figure files
\usepackage{amssymb}   % for math
\usepackage{dcolumn}% Align table columns on decimal point
\usepackage{bm}% bold math
\usepackage[colorlinks,citecolor=blue, linkcolor=blue,hyperindex,CJKbookmarks]{hyperref}
\begin{document}

\vspace{2.5cm}
\title{Classical-driving-enhanced parameter-estimation precision of a non-Markovian dissipative two-state system}
\author{Yan-Ling Li$^{1}$}
\author{Xing Xiao$^{2,3}$}
\altaffiliation{xiaoxing1121@gmail.com}
\author{Yao Yao$^{3}$}
\altaffiliation{yaoyao@csrc.ac.cn}
\affiliation{$^{1}$ School of Information Engineering, Jiangxi University of Science and Technology, Ganzhou 341000, China\\
$^{2}$College of Physics and Electronic Information,
Gannan Normal University, Ganzhou 341000, China\\
$^{3}$ Beijing Computational Science Research Center, Beijing 100084, China \\}

\begin{abstract}
The dynamics of quantum Fisher information (QFI) of the phase parameter in a driven two-state
system is studied within the framework of non-Markovian dissipative process. The influences of
memory effects, classical driving and detunings on the parameter-estimation precision are demonstrated by exactly solving the Hamiltonian under
rotating-wave approximation. In sharp contrast with the results obtained in
the presence of Markovian dissipation, we find that classical driving can
drastically enhance the QFI, namely, the precision of parameter estimation in the non-Markovian regime.
Moreover, the parameter-estimation precision may even
be preserved from the influence of surrounding
non-Markovian dissipation with the assistance of classical driving. Remarkably, we reveal that
the enhancement and preservation of QFI highly depend on the combination of classical
driving and non-Markovian effects. Finally, a phenomenological explanation of the underlying mechanism
is presented in detail via the quasimode theory.

\end{abstract}
\maketitle

\section{Introduction}
\label{intro}
Quantum metrology is a fast developing field of current research in both theoretical and experimental
physics\cite{giov06,giov11}. It is aimed to explore the capabilities of quantum systems that, when employed as probes
sensing physical parameters, allow to attain resolutions that are beyond the ability of classical protocols \cite{kay,helstrom}.
According to the quantum estimation theory, the ultimate achievable precision in parameter estimation scenarios is characterized by the quantum
Cram\'{e}r-Rao inequality \cite{braunstein}: $\delta\phi\geq1/\sqrt{NF}$, where $N$ denotes the number of measurement repetitions and $F$ is the QFI. Namely, the ultimate precision is inversely proportional to the square root of QFI.

Phase estimation plays a central role in practical quantum metrology, such as
optical interferometry \cite{dowling98,giov04} and atomic spectroscopy \cite{wineland92,boll96}, since most of tasks can be attributed to the problem of estimating the relative phase. Unfortunately, any realistic quantum system inevitably couples to
an uncontrollable environment which influences it in a non-negligible way \cite{breuer1}.
Then the issue of robustness of quantum metrological protocols against
various sources of decoherence has soon been raised \cite{alipour,tsang,benatti,yao,xiao14}, in particular, wondering whether such metrological schemes could be utilized not only to beat the standard quantum limit (SQL), but also to achieve the Heisenberg limit (HL) \cite{giov06}. In this context, it is a pivotal task to preserve the precision scaling under the environmental noises.

%
%According to the quantum estimation theory \cite{helstrom}, the ultimate achievable precision in parameter
%estimation scenarios is characterized by the quantum
%Cram\'{e}r-Rao inequality \cite{braunstein}.
%\begin{equation}
%\label{e1} \delta\phi\geq\frac{1}{\sqrt{NF}},
%\end{equation}
%where $N$ denotes the number of
%measurement repetitions and $F$ is the QFI \cite{holevo}.
%Namely, the ultimate precision is inversely proportional to the square root
%of QFI. Therefore, preserving the
%precision scaling under decoherence can be translated into the
%protection of QFI in the presence of environmental noise. Yet,
%the efficiency of protecting QFI is greatly dependent on the
%intrinsic properties of the environment coupled to the system.

The feasibility of preservation is greatly dependent on the intrinsic properties of the environment coupled to the system. According to the scale of correlation times, environments can be grouped into Markovian or non-Markovian types. The former
case possessing a small decoherence time during which correlations disappear, has been proven to be harmful to quantum metrology \cite{fujiwara,kolo,knysh11,escher,huelga,Demkowicz09,Genoni11,Demkowicz12,Vidrighin14}.
Even a very low noise level can completely
destroy the superiority of quantum metrological protocols and turn HL into SQL. However, non-Markovian environments which characterized by long correlation
times or structured spectral features would be more general in
many physical situations \cite{lambropoulos,breuer2,xiaoepjd,thorwart,sarovar}. The research of non-Markovian effects (also known as memory effects) is attracting extensive attentions due to key developments in the analysis,
understanding, and even simulation of nontrivial system-environment effects \cite{bellomo,fff,lu10,breuer3,liu,zhangwm,Matsuzaki,berrada13}.
In a seminal work of Chin \emph{et al} \cite{chin}, they first pointed that the non-Markovian effects can guarantee the advantage of quantum metrological strategies in the
presence of noise. This means non-Markovianity may serve as a new
resource for enhancing estimation tasks in open systems. The aforementioned discussions are restricted to consider the influence of
non-Markovian effects on the parameter estimation. However, to the best of our knowledge, few detailed investigations concerning the control of precision under environmental
noises are available at present.

Motivated by the above considerations, this study is to
discuss the role of classical driving in the precise
estimation of relative phase. To this end, the QFI is examined for a
driven two-level system in a zero-temperature non-Markovian reservoir.
Our results indicate that the precision can be drastically enhanced and even be completely preserved with the assistance of classical driving performed on the
qubit. Moreover, two factors for enhancing and preserving the precision are explored by
comparing with the results in Markovian case: the classical driving and
the memory effect of non-Markovian reservoir. The
memory effect provides the feasibility of the enhancement
while the classical driving provides a way to improve the precision.
Our results provide an
active way to suppress decoherence and enhance the parameter-estimation precision,
which is rather significant in quantum precision measurement and quantum metrology.

This paper is organized as follows. In Sec.~\ref{sec:2}, we
review the fundamental concept of QFI. In Sec.~\ref{sec:3},
the exact dynamics of a two-state system driven by classical fields and embedded in a
zero-temperature
non-Markovian environment is investigated. In Sec.~\ref{sec:4},
we show that the precision of parameter-estimation could be drastically enhanced
with the assistance of classical driving. The influences of other factors,
such as detunings and non-Markovian effects on the precision are discussed.
Next, we reveal the underlying physical
mechanism of the classical-driving-enhanced parameter-estimation precision via the quasimode theory
in Sec.~\ref{sec:5}. Finally,
Sec.~\ref{sec:6} gives a brief summary.

\section{Quantum Fisher Information}
\label{sec:2} The classical Fisher information originates from the
statistical inference, in which we are given a set probability
distributions $p(x_{i}|\phi)$ with measurement outcomes $\{x_i\}$.
 Here $\phi$ is an unknown parameter that we wish to determine and $\vec{X}=\{x_1, x_2,..., x_N\}$
 is an observable random variable. Without loss of generality, we assume
$\phi$ is a real parameter and the observable $\vec{X}$ is
discrete. The classical Fisher information is defined as
\begin{equation}
F_{\phi}=\sum_{i}p(x_{i}|\phi)\left[\frac{\partial\ln
p(x_{i}|\phi)}{\partial\phi}\right]^2, \label{e2}
\end{equation}
which characterizes the inverse variance of the asymptotic normality
of a maximum-likelihood estimator. Note that if the observable $X$
is continuous, the summation should be replaced by an integral.

Quantum Fisher information is formally generalized from the
classical one and is defined as
\begin{equation}
F_{\phi}=\textrm{Tr}(\rho_{\phi}
L^{2}_{\phi})=\textrm{Tr}[(\partial_{\phi} \rho_{\phi})L_{\phi}],
\label{e3}
\end{equation}
where $\mathcal {L}_{\theta}$ is the so-called symmetric logarithmic
derivative, which is defined by
$\partial_{\phi}\rho_{\phi}=(\mathcal
{L}_{\phi}\rho_{\phi}+\rho_{\phi}\mathcal {L}_{\phi})/2$ with
$\partial_{\phi}=\partial/\partial\phi$. By diagonalizing the matrix
as
$\rho_{\phi}=\Sigma_{n}\lambda_{n}|\psi_{n}\rangle\langle\psi_{n}|$,
one can rewritten the QFI as \cite{knysh11,liu13}
\begin{equation}
F_{\phi}=\sum_{n}\frac{(\partial_{\phi}\lambda_{n})^2}{\lambda_{n}}+\sum_{n}\lambda_{n}F_{\phi,n}
-\sum_{n\neq
m}\frac{8\lambda_{n}\lambda_{m}}{\lambda_n+\lambda_m}|\langle\psi_{n}|\partial_{\phi}\psi_{m}\rangle|^2,
\label{e4}
\end{equation}
where $\mathcal {F}_{\phi,n}$ is the QFI for pure state
$|\psi_{n}\rangle$ with the form
\begin{equation}
F_{\phi,n}=4[\langle\partial_{\phi}\psi_{n}|\partial_{\phi}\psi_{n}\rangle-|\langle\psi_{n}|\partial_{\phi}\psi_{n}\rangle|^2].
\label{e5}
\end{equation}
Note that Eq. (\ref{e4}) suggests the QFI of a non-full rank state
is only determined by the subset of $\{|\psi_{i}\rangle\}$ with
nonzero eigenvalues. Physically, the QFI can be divided into three
parts \cite{liu13,zhang13}. The first term is just the classical
Fisher information determined by the probability distribution; The
second term is a weighted average over the QFI for all the nonzero
eigenstates; The last term stemming from the mixture of pure states
reduces the QFI and hence the estimation precision below the
pure-state case.

\section{Model}
\label{sec:3}

We consider a two-level atom with Bohr frequency $\omega_{0}$ driven by a classical field of frequency $\omega_{L}$ \cite{haikka10,xiao10}.
The atom is embedded in a zero-temperature bosonic reservoir. Under the rotating-wave approximation,
the Hamiltonian of the system can be written as ($\hbar=1$)
\begin{equation}
\label{e7}
H=\frac{\omega_{0}}{2}\sigma_{z}+\sum_{k}\omega_{k}a_{k}^{\dagger}a_{k}+\left(\sum_{k}g_{k}a_{k}\sigma_{+}+\Omega
e^{-i\omega_{L}t}\sigma_{+}+h.c.\right),
\end{equation}
where $\sigma_{x,y,z}$ are the Pauli operators, $\sigma_{+}$ and $\sigma_{-}$
the atomic inversion operators, $a_{k}^{\dagger}$ and $a_{k}$
the creation and annihilation operators of the $k$th mode with
frequency $\omega_{k}$ of the reservoir and $g_{k}$ the coupling
constants between the atom and the reservoir.
The Rabi frequency $\Omega$, which has been assumed to be a real number, is taken to
be small compared to the atomic and laser frequencies $\Omega\ll\omega_{0},\omega_{L}$.

Since a unitary transformation does not change the eigenvalues of
the system, in the rotating reference frame through a unitary
transformation $U_{R}=e^{-i\omega_{L}\sigma_{z}t/2}$, the
Hamiltonian in equation (\ref{e7}) is equivalently transferred to an
effective Hamiltonian
\begin{equation}
\label{e8}
H_{e}=\frac{\Delta}{2}\sigma_{z}+\Omega\sigma_{x}+\sum_{k}\omega_{k}a_{k}^{\dagger}a_{k}+\left(\sum_{k}g_{k}a_{k}\sigma_{+}e^{i\omega_{L}t}+h.c.\right),
\end{equation}
where $\Delta=|\omega_{0}-\omega_{L}|$. Note that the first two terms on the
right side can be diagonalized in the new dressed bases
\begin{eqnarray}
\label{e9}
|E\rangle&=&\cos\frac{\eta}{2}|e\rangle+\sin\frac{\eta}{2}|g\rangle,\nonumber\\
|G\rangle&=&-\sin\frac{\eta}{2}|e\rangle+\cos\frac{\eta}{2}|g\rangle,
\end{eqnarray}
with $\eta=\tan^{-1}(2\Omega/\Delta)$. Then in the dressed-state bases, the effective Hamiltonian can be
rewritten as
\begin{equation}
\label{e10}
H_{e}'=\frac{\omega_{D}}{2}\rho_{z}+\sum_{k}\omega_{k}a_{k}^{\dagger}a_{k}+\cos^2\frac{\eta}{2}\sum_{k}(g_{k}e^{i\omega_{L}t}a_{k}\rho_{+}+h.c.),
\end{equation}
where $\omega_{D}=\sqrt{\Delta^2+4\Omega^2}$ is the dressed
frequency. The new inversion operator $\rho_{z}$ is given by
$\rho_{z}=|E\rangle\langle E|-|G\rangle\langle G|$ and the new
raising operator $\rho_{+}$ is defined as $\rho_{+}=|E\rangle\langle
G|$. Here, the terms $a_{k}\rho_{z}e^{i\omega_{L}t}$,
$a_{k}\rho_{-}e^{i\omega_{L}t}$ and their complex conjugates have
been neglected by using the usual rotating-wave approximation.

We consider the situation of no more than one
excitation in the whole system, so the subspace spanned in the dressed bases is given by: $
|\psi_{0}\rangle=|G\rangle_{S}\otimes|0\rangle_{R},
|\psi_{1}\rangle=|E\rangle_{S}\otimes|0\rangle_{R},
|\psi_{k}\rangle=|G\rangle_{S}\otimes|1_{k}\rangle_{R},$ where
$|1_{k}\rangle_{R}$ indicates that there is a photon in  the
\emph{k}th mode of the reservoir. Assuming the environment is initially
prepared in the vacuum state, then
it follows that any initial state
of the form
\begin{equation}
\label{e11}
|\Psi(0)\rangle=c_{0}|\psi_{0}\rangle+c_{1}(0)|\psi_{1}\rangle,
\end{equation}
evolves after time \emph{t} into the state
\begin{equation}
\label{e12}
|\Psi(t)\rangle=c_{0}|\psi_{0}\rangle+c_{1}(t)|\psi_{1}\rangle+\sum_{k}c_{k}(t)|\psi_{k}\rangle.
\end{equation}
Note that the amplitudes $c_{1}(t)$ and $c_{k}(t)$ depend on time,
while the amplitude $c_0$ is constant in time because
$H_{e}^{'}|\psi_{0}\rangle=0$. By solving the Schr\"{o}dinger
equation, we could obtain a closed integro-differential equation for
$c_{1}(t)$
\begin{equation}
\label{e13}
\dot{c}_{1}(t)=-\cos^4\frac{\eta}{2}\int_{0}^{t}dt_{1}f(t-t_1)c_{1}(t_{1}).
\end{equation}
The kernel $f(t-t_1)$ is given by a certain two-point correlation
function of the reservoir \cite{breuer1}.
\begin{equation}
\label{e14}
f(t-t_1)=\int d\omega J(\omega
)\exp [i(\omega_{D} +\omega_{L}-\omega)(t-t_{1})].
\end{equation}
In above, we have used the limitation of a continuum of reservoir modes
$\sum_{k}|g_{k}|^2\rightarrow\int J(\omega)d\omega$, where $J(\omega)$
is the spectral density function, characterizing
the reservoir spectrum. Until now, our result is valid for an environment with a generic
spectral density since no restrictive hypothesis is made on the
environment.

In order to study in more detail the QFI in a non-Markovian
environment, we need to specify the spectral density of the
reservoir. We focus on the Lorentzian spectral density of the form
\begin{equation}
\label{e15}
 J(\omega)=\frac{1}{2\pi}\cdot\frac{\gamma_{0}\lambda^{2}}{(\omega_{0}-\omega-\delta)^{2}+\lambda^{2}},
\end{equation}
where $\delta=\omega_0-\omega_{c}$ is the detuning between atomic frequency $\omega_0$ and
the center frequency of the structured environment $\omega_{c}$. The parameter
$\lambda$ defines the spectral width and $\gamma_0$
is related to the decay of the excited state of the
atom in the Markovian limit \cite{spohn}. Usually, there are two regimes
\cite{dalton}: weak-coupling regime ($\gamma_0<\lambda/2$), where the
behavior of the system is Markovian,
and strong-coupling regime ($\gamma_0>\lambda/2$), where non-Markovian dynamics occurs.
Note that Eq.(\ref{e15}) is one of the most studied spectrums
of bosonic environments since it leads to an electromagnetic field
inside an imperfect cavity supporting the mode $\omega_0$, which is
typical of dissipative systems in several physical contexts. In this
model, the reservoir correlation function is given by an exponential
form
\begin{equation}
\label{e16}
f(t-t_1)=\frac{\gamma_{0}\lambda}{2}\exp\left[-M(t-t_{1})\right],
\end{equation}
where with $M=\lambda+i\Delta-i\delta-i\omega_{D}$.
Then the probability amplitude can be easily calculated
$c_{1}(t)=c_{1}(0)\xi(t)$, where $\xi(t)$ is expressed
as
\begin{equation}
\label{e17}
\xi(t)=e^{-Mt/2}[\cosh(\frac{Kt}{4})+\frac{2M}{K}\sinh(\frac{Kt}{4})],
\end{equation}
with $K=\sqrt{4M^2-2\gamma_{0}\lambda(1+\cos\eta)^2}$. Briefly, the total
evolution could be described by the following state map
\begin{eqnarray}
\label{e18}
|G\rangle_{S}\otimes|0\rangle_{R}&&\rightarrow|G\rangle_{S}\otimes|0\rangle_{R},\\
\label{e19}
|E\rangle_{S}\otimes|0\rangle_{R}&&\rightarrow\xi(t)|E\rangle_{S}\otimes|0\rangle_{R}+\sqrt{1-\xi^{2}(t)}|G\rangle_{S}\otimes|1_{k}\rangle_{R}.\nonumber
\end{eqnarray}

\section{Classical-driving-enhanced parameter-estimation precision}
\label{sec:4}

With above state map in mind, we assume that the qubit is initially
in the state
\begin{equation}
\label{e20} |\psi\rangle_{a} = \cos \frac{\theta}{2}|0\rangle +
e^{i\phi} \sin \frac{\theta}{2}|1\rangle,
\end{equation}
and the reservoir is in the vacuum state, then the state of the
total system is
\begin{eqnarray}
\label{e21}
|\Psi(t)\rangle&=&\big[c_{0}\cos\frac{\eta}{2}+c_{1}\xi(t)\sin\frac{\eta}{2}\big]|0\rangle_{a}|0\rangle_{R}\nonumber\\
&&+\big[-c_{0}\sin\frac{\eta}{2}+c_{1}\xi(t)\sin\frac{\eta}{2}\big]|1\rangle_{a}|0\rangle_{R}\nonumber\\
&&+\big[d(t)\cos \frac{\eta}{2}|0\rangle_{a}-d(t)\sin \frac{\eta}{2}|1\rangle_{a}\big]|1_{k}\rangle_{R},
\end{eqnarray}
where
$c_{0}=\cos\frac{\theta}{2}\cos\frac{\eta}{2}-e^{i\phi}\sin\frac{\theta}{2}\sin\frac{\eta}{2}$,
$c_{1}=\cos\frac{\theta}{2}\sin\frac{\eta}{2}+e^{i\phi}\sin\frac{\theta}{2}\cos\frac{\eta}{2}$,
$d(t)=\sqrt{1-|c_{1}\xi(t)|^{2}-|c_{0}|^{2}}$. The reduced density
matrix of the atomic qubit can be given by tracing the reservoir's
degrees of freedom
\begin{equation}
\label{e22}
\rho(t) =
\rho_{00}|0\rangle\langle0|+\rho_{01}|0\rangle\langle1|+\rho_{10}|1\rangle\langle0|+\rho_{11}|1\rangle\langle1|,
\end{equation}
where
\begin{eqnarray}
\label{e23}
\rho_{00}&=&\cos^{2}\frac{\eta}{2}-|c_{1}\xi(t)|^{2}\cos\eta+\frac{1}{2}[c_{0}c^{*}_{1}\xi^{*}(t)+c^{*}_{0}c_{1}\xi(t)]\sin\eta,\nonumber\\
\rho_{01}&=&[-\frac{1}{2}+|c_{1}\xi(t)|^{2}]\sin\eta+c_{0}c^{*}_{1}\xi^{*}(t)\cos^{2}\frac{\eta}{2}-c^{*}_{0}c_{1}\xi(t)\sin^{2}\frac{\eta}{2},\nonumber\\
\rho_{10}&=&[-\frac{1}{2}+|c_{1}\xi(t)|^{2}]\sin\eta-c_{0}c^{*}_{1}\xi^{*}(t)\sin^{2}\frac{\eta}{2}+c^{*}_{0}c_{1}\xi(t)\cos^{2}\frac{\eta}{2}\nonumber\\
\rho_{11}&=&\sin^{2}\frac{\eta}{2}+|c_{1}\xi(t)|^{2}\cos\eta-\frac{1}{2}[c_{0}c^{*}_{1}\xi^{*}(t)+c^{*}_{0}c_{1}\xi(t)]\sin\eta.
\end{eqnarray}

For the single qubit state, according to Eq. (\ref{e4}), an
explicitly expression of QFI could be obtained. In the Bloch sphere
representation, any qubit state can be written as
\begin{equation}
\rho = \frac{1}{2}(1+\vec{W}\cdot\hat{\sigma}),
\end{equation}
where $\vec{W}=(W_{x}, W_{y}, W_{z})^{T}$ is the real Bloch vector
and $\hat{\sigma}=(\hat{\sigma}_{x}, \hat{\sigma}_{y},
\hat{\sigma}_{z})$ denotes the Pauli matrices. Therefore, for the
single qubit state, $F_{\phi}$ can be represented as follows
\cite{zhong13}

\begin{eqnarray}
F_{\phi}=\left\{\begin{array}{cc}
|\partial_{\phi} \vec{W}|^{2}+\frac{\vec{W}\cdot\partial_{\phi}\vec{W}}{1-|\vec{W}|^{2}}, & \mbox{ if } \, |\vec{W}|<1,\\
|\partial_{\phi}\vec{W}|^{2}, & \mbox{ if } \, |\vec{W}|=1.
\end{array}\right.
\label{e24}
\end{eqnarray}

Substituting the Bloch vector components $W_{x}=\rho_{01}+\rho_{10},
W_{y}=i(\rho_{01}-\rho_{10}), W_{z}=2\rho_{00}-1$ into Eq. (\ref{e4}), then
the dynamics of the QFI of parameter $\phi$ can be obtained exactly
by Eq.(\ref{e24}). However, the explicit expression is too
complicated to present in the text. Nevertheless, numerical results
indicate that the dynamics of QFI in non-Markovian reservoir by
classical driving show interesting properties.

\begin{figure}
  \includegraphics[width=0.5\textwidth]{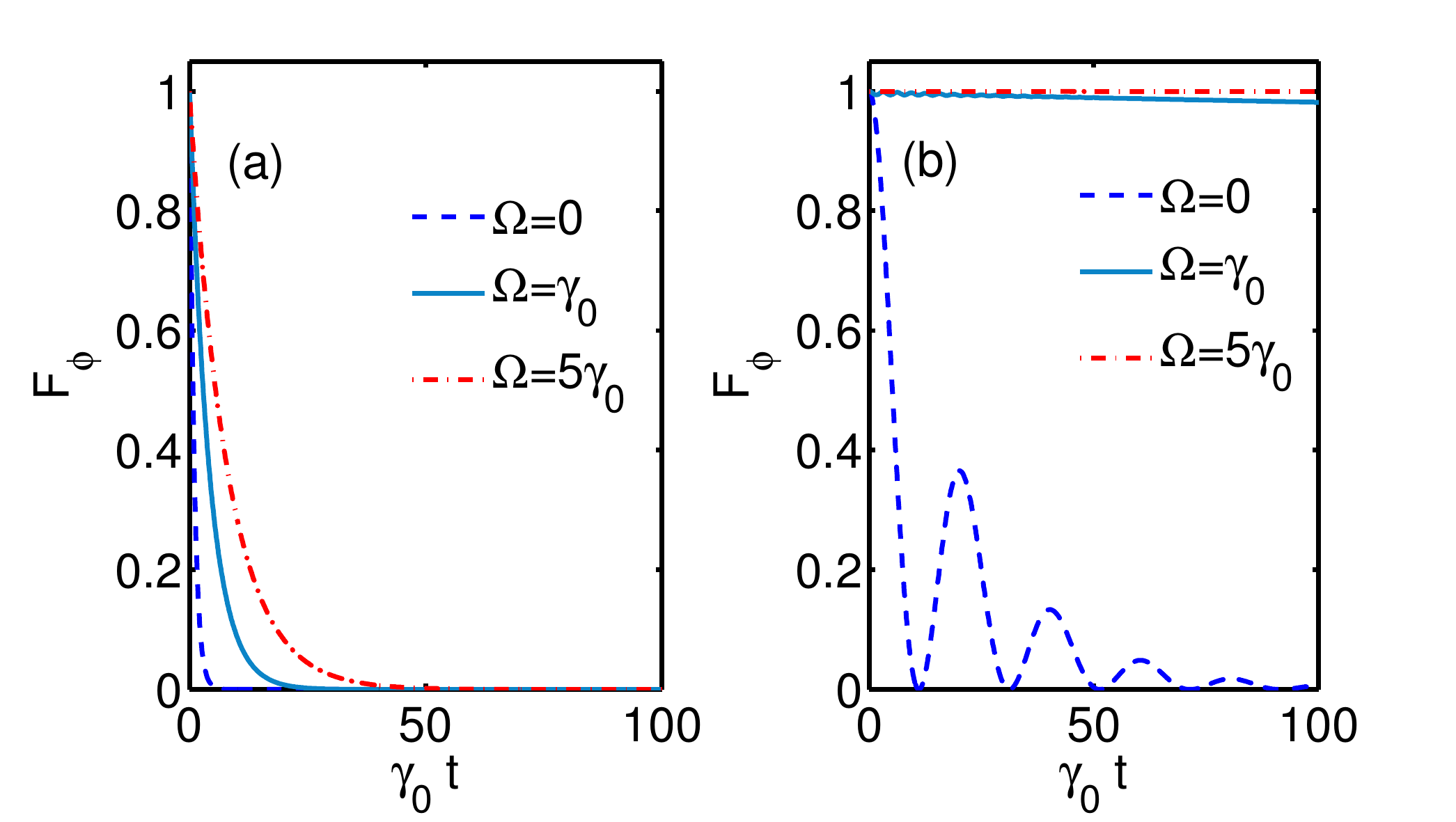}
\caption{(color online) Quantum Fisher information $F_{\phi}$ as a
function of $\gamma_{0}t$ under the classical driving. The other
parameters are $\theta=\pi/2$, $\Delta=0$ and $\delta=0$. (a) In the
Markovian regime $\lambda=10\gamma_{0}$. (b) In the non-Markovian
regime $\lambda=0.05\gamma_{0}$.}
\label{fig1}       % Give a unique label
\end{figure}

In order to observe the effect of classical driving on the
estimation precision clearly, in Fig.~\ref{fig1}, we show the
dynamics of the QFI ($F_{\phi}$) with respect to different strength
of classical driving. In order to make results comparable, we have
plotted the results in both Markovian and non-Markovian regimes. It
is clearly shown that the classical driving may slightly retard the QFI loss
during the time evolution in the Markovian regime, but the
$F_{\phi}$ still decays rapidly to zero. The decay of $F_{\phi}$
reflects that the estimation of parameter $\phi$ becomes more
inaccurate in this situation. In contrast, the behaviors of QFI are
more complicated and interesting in the non-Markovian regime. As
shown in Fig.~\ref{fig1}(b), the QFI experiences damped oscillations
in the absence of classical driving. The reason is that the QFI flows
back and forth between the system and its non-Markovian reservoir
due to the memory effect (a long correlation time of the reservoir)
\cite{lu10,berrada13}. However, it is remarkable that in the
non-Markovian regime the classical driving plays an important
role in the dynamics of QFI. We find the classical driving can
dramatically protect the QFI from the influence of non-Markovian
noises and enhance the precision of parameter estimation. The
stronger the classical driving is, the higher is the estimation
precision. These results can be understood as follows. On the one
hand, we know that the oscillations of the atomic inversion
represent the exchanging of energy between the atom and the field.
When the Rabi frequency becomes larger, the energy exchange will
become more rapid \cite{scully}. On the other hand, when the Rabi frequency
increases, the effective coupling between the qubit and the
reservoir decreases that suppresses the information exchange.
Therefore the outflow of the information from the qubit is suppressed,
i.e., the decay of QFI slows down. Due to the memory
effects, this phenomenon is more evidently in the non-Markovian
regime than Markovian regime. However, what's the relation between
the QFI preservation, non-Markovian effects and the classical
driving?
\begin{figure}
  \includegraphics[width=0.4\textwidth]{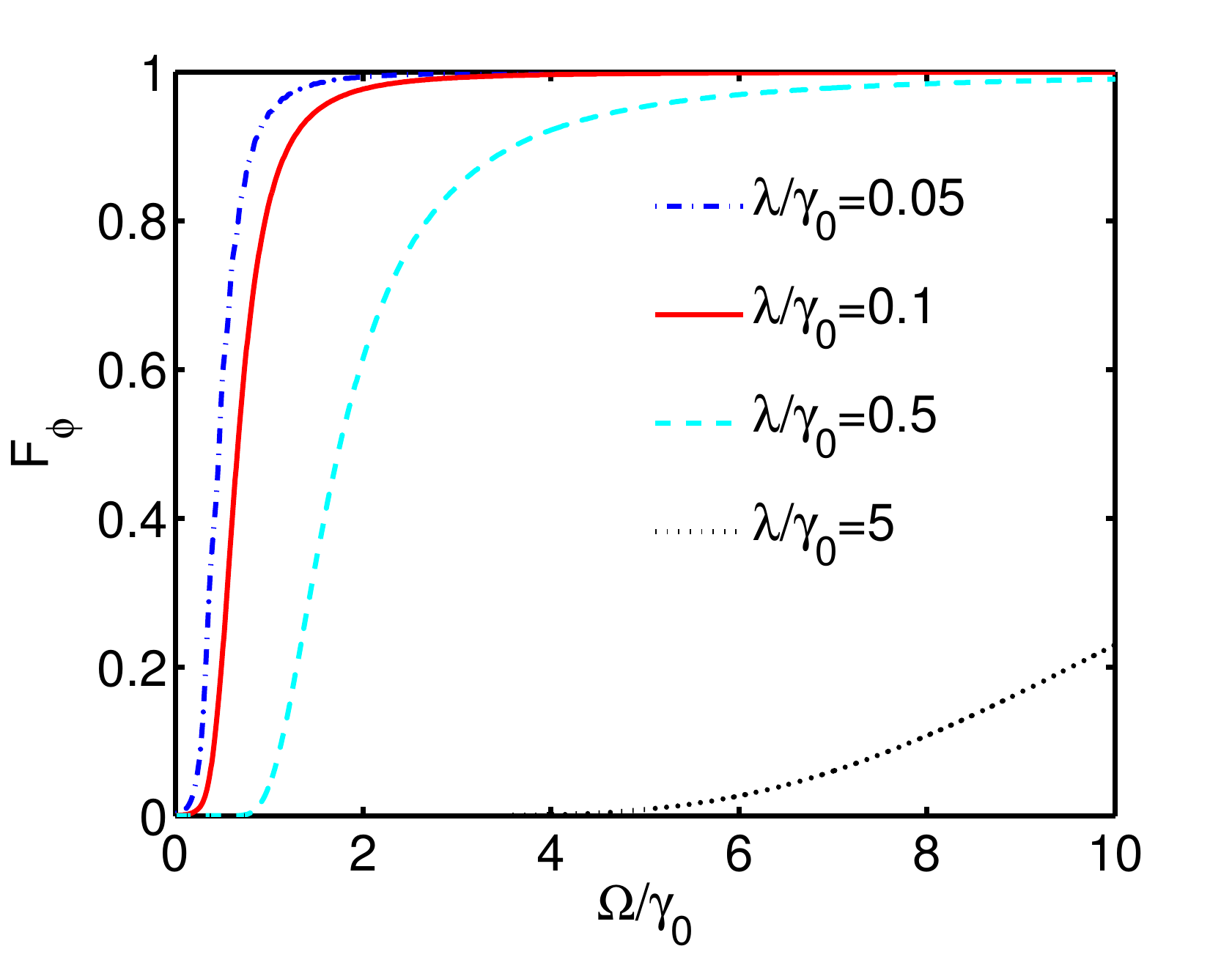}
\caption{(color online) Asymptotic behaviors of QFI as a function of
the dimensionless quantity $\Omega/\gamma_0$ with $\theta=\pi/2$,
$\Delta=0$, $\delta=0$ and $\gamma_{0}t=50$. Here,
$\lambda=5\gamma_0$ corresponds to the Markovian reservoir, while
$\lambda=0.05\gamma_0$, $\lambda=0.1\gamma_0$, $\lambda=0.5\gamma_0$
correspond to different non-Markovian reservoirs, respectively.}
\label{fig2}       % Give a unique label
\end{figure}

To get a better understanding of the effects of classical
driving and non-Markovian characteristics on the QFI
preservation, we plot Fig.~\ref{fig2} to show the asymptotic
behaviors of QFI as a function of the dimensionless quantity
$\Omega/\gamma_0$ for different values of $\lambda/\gamma_0$ with
$\gamma_{0}t=50$. We can find that the correlation time of the
reservoir significantly affects the precision of parameter
estimation. The amplitude of the QFI increases with the decrease of
the $\lambda/\gamma_{0}$. This is because the smaller the value of
$\lambda/\gamma_{0}$, the stronger the non-Markovian effects. And
more information can be feed back to the system. Then the precision
of parameter estimation can be improved by the enhancement of the
non-Markovian effects. On the other hand, if the strength of non-Markovian effects
is weak, for instance $\lambda/\gamma_0=0.5$, then one can alternatively enhance the 
QFI by increasing the classical driving parameter
$\Omega$. Remarkably, the efficiency of QFI preservation
approximates to 100\% when $\Omega=10\gamma_{0}$ in the
non-Markovian environments. While in the Markovian regime
($\lambda/\gamma_0=5$), only a small amount of QFI could be
preserved. These results indicate that the enhancement of QFI
greatly benefits from the combination of classical driving and
non-Markovian effects.

\begin{figure}
  \includegraphics[width=0.4\textwidth]{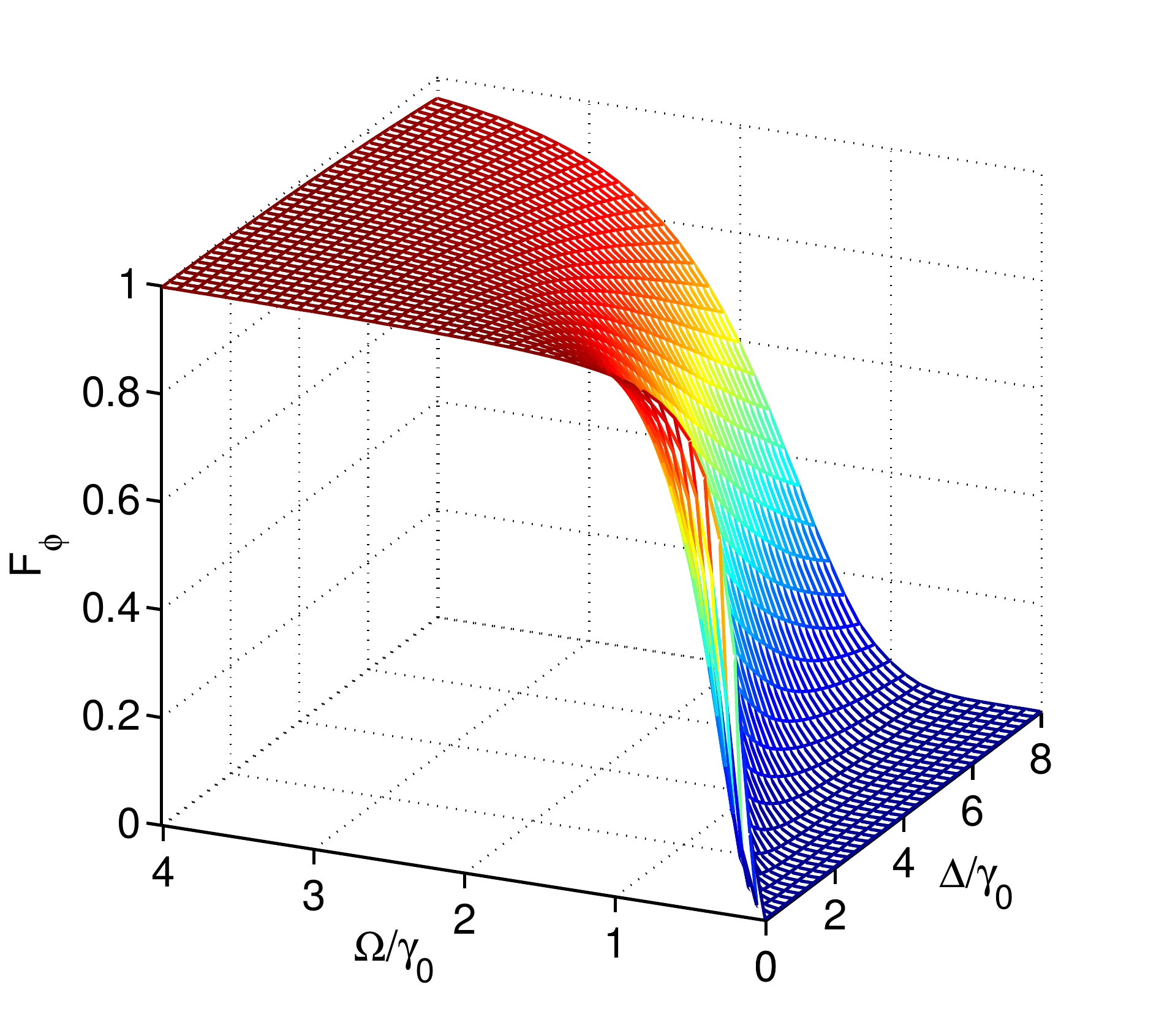}
\caption{(color online) QFI as a function of detuning $\Delta/\gamma_0$ and
the strength of classical driving $\Omega/\gamma_0$ in non-Markovian regime ($\lambda=0.1\gamma_0$).
The other parameters are
$\theta=\pi/2$, $\gamma_{0}t=50$ and $\delta=0$.}
\label{fig3}       % Give a unique label
\end{figure}

In above, only the resonantly driving case is discussed, but the
detuning (i.e. $\Delta \neq 0$) case would be more reasonable
considering the frequency shift induced by the interaction with the
environment. $F_{\phi}$ as a function of dimensionless quantity
$\Delta/\gamma_0$ and $\Omega/\gamma_0$ with $\gamma_{0}t=50$ is
plotted in Fig.~\ref{fig3}. It follows from the numerical analysis
that the detuning $\Delta$ has an adverse influence upon the precision
of parameter estimation. This result is to be expected since the
large detuning make the coupling between the qubit and the driving
field weaker. Fortunately, this negative influence can be suppressed
by increasing the the strength of classical driving.

\begin{figure}
  \includegraphics[width=0.4\textwidth]{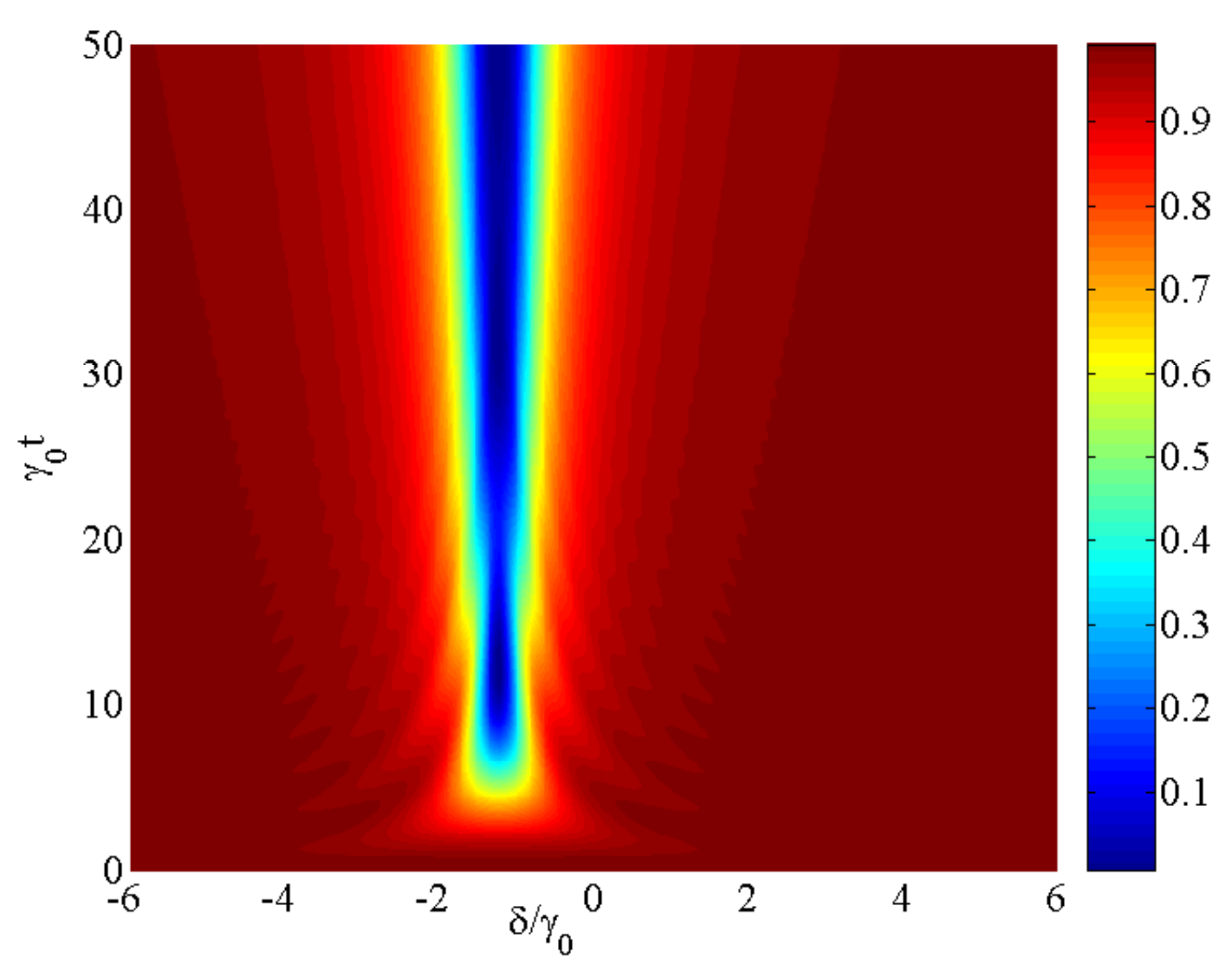}
\caption{(color online) The contour plot of QFI as a function of $\gamma_{0} t$ and qubit-reservoir detunings $\delta/\gamma_0$. The other parameters are $\lambda=0.1\gamma_{0}$,
$\theta=\pi/2$, $\Omega=\gamma_0$ and $\Delta=\gamma_0$.}
\label{fig4}       % Give a unique label
\end{figure}

Another factor on the precision of parameter estimation is the
detuning $\delta=\omega_0-\omega_c$ between the qubit frequency
$\omega_{0}$ and the center frequency of the structured reservoir
$\omega_c$. In Ref.\cite{berrada13}, the author has shown
that the enhancement of the QFI may occur by adjusting the
reservoir-qubit detuning. However, we should point out that the result
is a little different in our model. It is noted that the enhancement
of QFI depends on the absolute values of $\delta$ in Ref. \cite{berrada13}, namely, both positive and negative detunings would
enhance the QFI equally. While in our model, considering the presence of
classical driving, the symmetry of $F_{\phi}$ (with respect to $\delta=0$) is broken and positive $\delta$ 
outperforms the negative one (with the same absolute value) for the enhancement of precision, as shown in Fig.~\ref{fig4}. In addition, we note that
negative $\delta$ first fails to improve the precision in a small region $[-(\sqrt{\Delta^2+4\Omega^2}-\Delta),0]$,
but regains the ability when $\delta<-(\sqrt{\Delta^2+4\Omega^2}-\Delta)$. This phenomenon could be understood
clearer in Sec. \ref{sec:5}.

In a nutshell, from Figs. 1-4, we note some
features as follows: (1) The increasing of Rabi frequency $\Omega$,
non-Markovian effects $\gamma_0/\lambda$ and the positive detuning
$\delta$ can effectively enhance the parameter-estimation precision.
(2) This drastic enhancement occurs only when the conditions of
the non-Markovian effects and the classical driving are satisfied
simultaneously. These results provide a good method to enhance the
precision of phase estimation in open quantum systems and would be
benefit for quantum metrology.

\section{Physical interpretation}
\label{sec:5}

In order to construct a more intuitive physical insight into the
phenomenon of classical-driving-enhanced precision in non-Markovian environment, we utilize the quasimode
Hamiltonian \cite{lang,barnett,dalton,dalton2} in the following analysis. For the Hamiltonian described by Eq. (\ref{e10}), we can
observe it in another rotating reference frame
$U=\exp[i\omega_{L}\rho_{z}t/2]$, then an effective Hamiltonian can be obtained as
\begin{equation}
H_{e}''=\frac{\omega_{0}'}{2}\rho_{z}+\sum_{k}\omega_{k}a_{k}^{\dagger}a_{k}+\sum_{k}(g_{k}'a_{k}\rho_{+}+h.c.),
\end{equation}
where $\omega_{0}'=\omega_0+\sqrt{\Delta^2+4\Omega^2}-\Delta$,
$g_{k}'=\cos^{2}(\eta/2)g_{k}$. The effective Hamiltonian $H_{e}''$ is an
exact unbiased spin-boson model. A noteworthy feature is that the
basis states have been changed to $\{|E\rangle,|G\rangle\}$ when the
atom coupled with the structured reservoir with the assistance of
the classical driving. Assuming the spectral function of the
reservoir is still Lorentzian, the corresponding quasimode
Hamiltonian can be given by
\begin{equation}
H_{quasi}=H_{0}'+H_{memory}+H_{dis},
\end{equation}
with
\begin{eqnarray}
&&H_{0}'=\frac{1}{2}\omega_{0}'\rho_{z}+\omega _{c}D^{\dagger }D+\int\nu C^{\dagger}(\nu)C(\nu)d\nu,\\
&&H_{memory}=\sqrt{\gamma_{0}\lambda/2}(\rho_{+}D+\rho_{-}D^{\dagger}),\\
&&H_{dis}=(\lambda/\pi)^{\frac{1}{2}}\int d\nu[D^{\dagger}C(\nu)+DC^{\dagger}(\nu)],
\end{eqnarray}
where $C^{\dagger}(\nu)$ and $C(\nu)$ are the creation and
annihilation operators of the continuum quasimode of frequency
$\nu$. $D^{\dagger}$ and $D$ are the creation and annihilation
operators of the discrete quasimode.

\begin{figure}
  \includegraphics[width=0.5\textwidth]{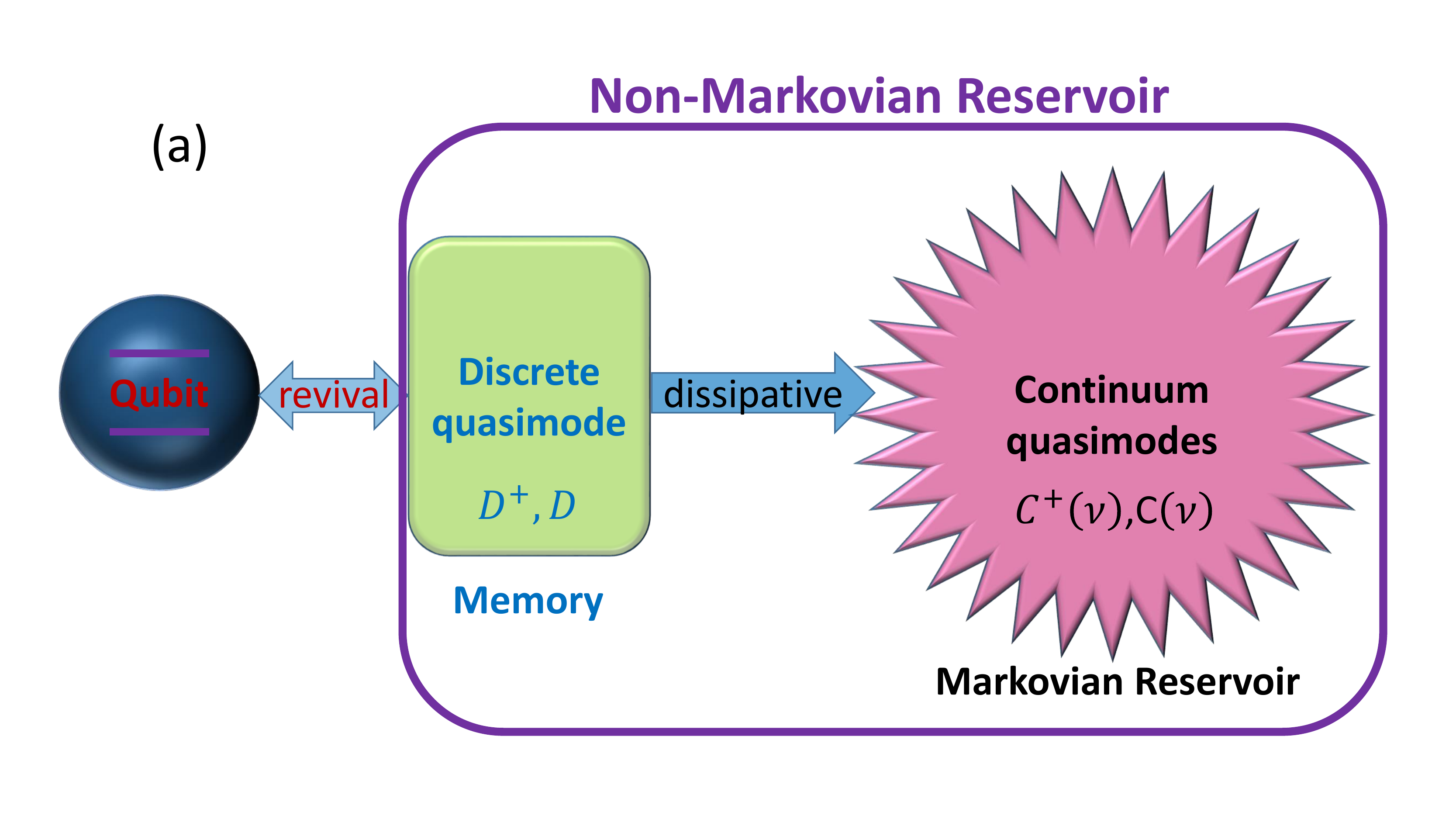}\\
  \includegraphics[width=0.4\textwidth]{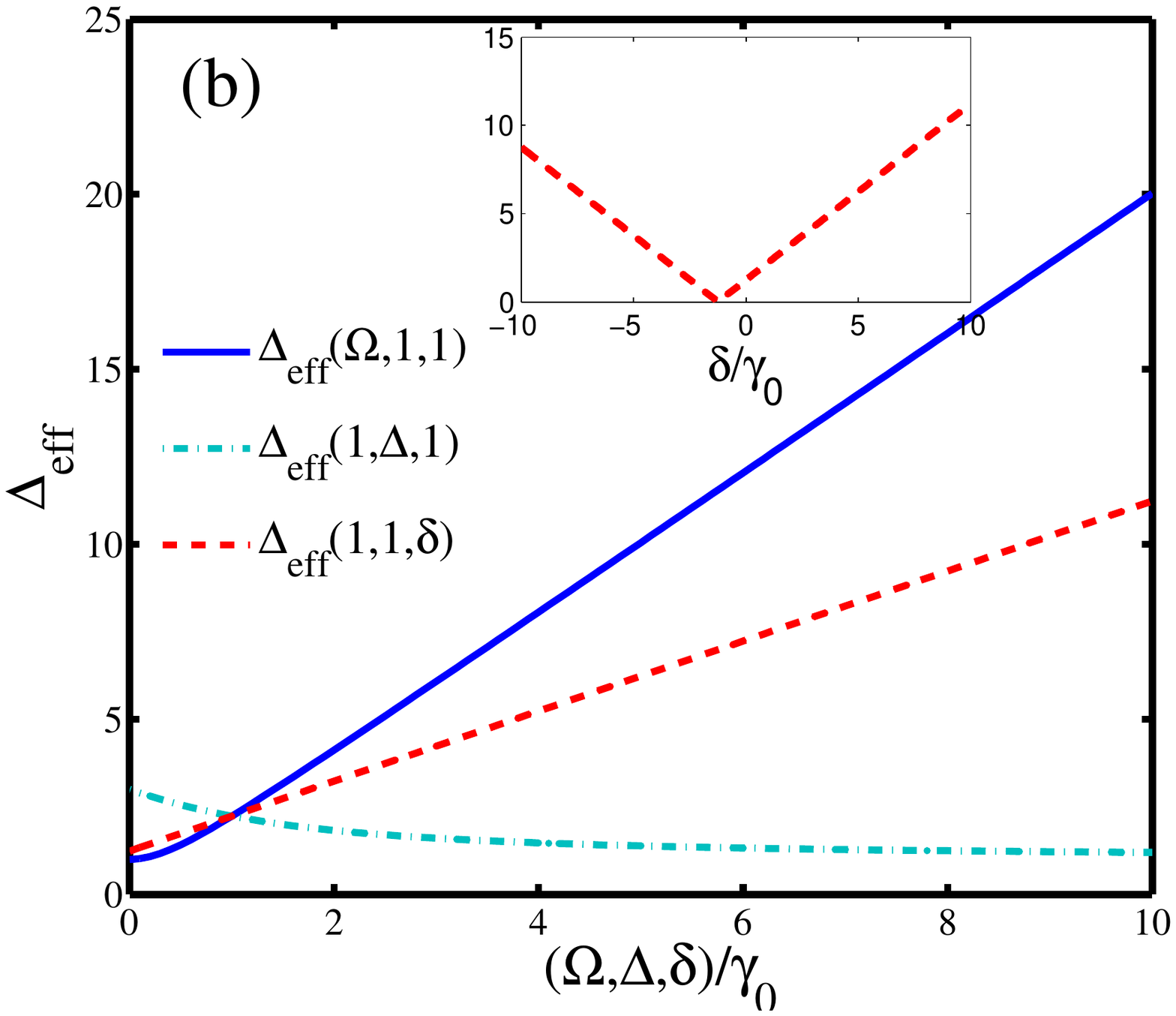}
\caption{(color online) (a) Illustration of a two-level system coupled
to a non-Markovian reservoir: the quasimode picture. In this
quasimode picture, the non-Markovian reservoir is divided into two
parts: memory (discrete quasimode) and Markovian reservoir
(continuum quasimodes). The two-level system is coupled to a
discrete quasimode, which is in turn coupled to external continuum
quasimodes. (b) The effective detuning $\Delta_{eff}(\Omega, \Delta, \delta)$ as a function of
  $\Omega$, $\Delta$ and $\delta$. The inset figure shows that negative $\delta$ first reduces the effective detuning $\Delta_{eff}$ in a small region and then increases it when $\delta<-(\sqrt{\Delta^2+4\Omega^2}-\Delta)$.}
\label{fig5}       % Give a unique
\end{figure}

We shall first exploit the physical mechanism of the QFI dynamics
without classical driving. From the quasimode Hamiltonian, we
find that the system only couples to a discrete quasimode, which
interacts with a set of continuum quasimodes, as shown in
Fig.~\ref{fig5}(a) Note that the coupling strength between the
discrete quasimode and continuum quasimodes just depends on the
width of the spectral density $\lambda$. The discrete quasimode
functions as a memory between the system and the dissipative environment.
The behavior of the new
system composed by the qubit and the discrete quasimode is
completely Markovian \cite{dalton}. Since the qubit interacts only with the
discrete quasimode, while the dissipative process happens only in
the interaction between the discrete mode and the continuum modes.
Therefore, in the Markovian regime ($\lambda\gg\gamma_0$), the
dynamics of QFI is dominated by the dissipative interaction
$H_{dis}$, and the QFI decays exponentially and vanishes
only asymptotically, as depicted in Fig.~\ref{fig1}(a). Whereas in
the non-Markovian regime $\lambda\ll\gamma_0$, the dissipative
interaction is suppressed and the information could exchange back
and forth between the system and discrete quasimode before it is
completely dissipated into the continuum quasimodes. So the QFI
vanishes with a damping of its revival amplitude in the
non-Markovian regime, as shown in Fig.~\ref{fig1}(b).

Next, we demonstrate how the classical driving affect the
dynamics of QFI. In the quasimode Hamiltonian, the memory part is
also a Jaynes-Cummings model, and the effective detuning between the
qubit and discrete quasimode is
\begin{equation}
\Delta_{eff}(\Omega, \Delta,
\delta)=|\sqrt{\Delta^2+4\Omega^2}-\Delta+\delta|,
\end{equation}
where $\Delta=|\omega_{0}-\omega_{L}|$, $\delta=\omega_{0}-\omega_{c}$
and Rabi frequency $\Omega$. It is well known that for the
Jaynes-Cummings model, the larger the absolute value of detuning, the weaker coupling
between the qubit and the mode of the field \cite{scully}. Then the information
transfer between qubit and field slows down. With this conclusion in
mind, the results in Sec. \ref{sec:4} could be explained clearly via
the effective detuning $\Delta_{eff}$. As shown in
Fig.~\ref{fig5}(b), $\Delta_{eff}$ is a monotonic increasing
function of $\Omega$ and positive $\delta$, while it is a decreasing
function of $\Delta$. Therefore, increasing $\Omega$ and positive
$\delta$ can effectively enhance the precision of parameter
estimation, whereas increasing $\Delta$ reduces the precision. That
is why increasing $\Omega$ and positive $\delta$ are beneficial to
the enhancement of precision, while increasing $\Delta$ is harmful. However, due to the
presence of classical driving, negative $\delta$ first reduces the effective detuning $\Delta_{eff}$ in a small region and then increases it when $\delta<-(\sqrt{\Delta^2+4\Omega^2}-\Delta)$. The sudden change point $\delta=-(\sqrt{\Delta^2+4\Omega^2}-\Delta)$, i.e., $\Delta_{eff}=0$, indicates
that the enhancement of precision provided by classical driving has been completely neutralized by the negative $\delta$.

We should also note that if the total
decoherence process is dominated by the dissipative part of the
quasimode Hamiltonian (i.e., in the Markovian regime), the
increasing of effective detuning $\Delta_{eff}$ just prolong the
decay slightly. Only when the dissipative channel is greatly
suppressed (i.e., in the non-Markovian regime), the increasing of
effective detuning $\Delta_{eff}$ can drastically enhance the
parameter-estimation precision. Namely, the enhancement is based on
the combination of large effective detuning between the system and
the discrete quasimode and the strong non-Markovian effects.

\section{Conclusions}
\label{sec:6}

In summary, we have investigated the parameter-estimation precision of a driven
two-state system coupled to a bosonic environment at zero temperature.
The precision is just slightly improved,
and still decays asymptotically
to zero in the Markovian environment under the classical driving. However, it could be greatly enhanced
and preserves a quite long time
in non-Markovian environment with the assistance of classical driving. We also find that
increasing the Rabi frequency or the
degree of the non-Markovian effects can make further improvement on the precision, while non-resonant driving
will reduce the precision. Moreover,
we should emphasize that the drastic enhancement is based on the combination
of classical driving and non-Markovian effects.
The above results provide an
active method to combat the influence of decoherence on quantum
metrology. Finally, according to the quasimode theory,
an intuitive physical interpretation has
been provided about the precision enhancement under the classical driving.

\begin{acknowledgements}
This work is supported by the Funds of the National Natural Science
Foundation of China under Grant Nos. 11247006, 11365011, 11447118
and the China Postdoctoral Science Foundation under Grant No. 2014M550598.
\end{acknowledgements}

%\begin{acknowledgements}
%If you'd like to thank anyone, place your comments here
%and remove the percent signs.
%\end{acknowledgements}

% BibTeX users please use one of
%\bibliographystyle{spbasic}      % basic style, author-year citations
%\bibliographystyle{spmpsci}      % mathematics and physical sciences
%\bibliographystyle{spphys}       % APS-like style for physics
%\bibliography{}   % name your BibTeX data base

% Non-BibTeX users please use

\end{document}